\documentclass[reprint, twocolumn, pra, a4paper, amssyb, amsmath, graphicx, superscriptaddress]{revtex4-2} 

\usepackage{changes}
\usepackage{hyperref}
\usepackage{comment}
\usepackage{textcmds}

\begin{document}

\title{Optimization of Richardson extrapolation for quantum error mitigation}
\author{Michael Krebsbach}
\email{mkrebsbach@web.de}
\affiliation{Institute for Theoretical Physics and Astrophysics, University of W\"urzburg, D-97074 W\"urzburg, Germany}
\author{Bj\"orn Trauzettel}
\affiliation{Institute for Theoretical Physics and Astrophysics, University of W\"urzburg, D-97074 W\"urzburg, Germany}
\affiliation{W\"urzburg-Dresden Cluster of Excellence ct.qmat, Germany}
\author{Alessio Calzona}
\affiliation{Institute for Theoretical Physics and Astrophysics, University of W\"urzburg, D-97074 W\"urzburg, Germany}
\affiliation{W\"urzburg-Dresden Cluster of Excellence ct.qmat, Germany}
\date{\today}
	
\begin{abstract}
Quantum error mitigation is a key concept for the development of practical applications based on current noisy intermediate scale quantum (NISQ) devices. One of the most promising methods is Richardson extrapolation to the zero noise limit. While its main idea is rather simple, the full potential of Richardson extrapolation has not been completely uncovered yet. We give an in-depth analysis of the relevant parameters of Richardson extrapolation and propose an optimized protocol for its implementation. 
This protocol allows for a precise control of the increase in statistical uncertainty and lays the foundation for a significant improvement of the mitigation performance achieved by increasing the number of nodes. Furthermore, we present a novel set of nodes that, on average, outperforms the linear, exponential or Chebyshev nodes frequently used for Richardson extrapolation without requiring any additional resources. 
\end{abstract}

\maketitle

One of the main challenges for the development of quantum computers (QCs) is their high sensitivity to errors and noise. While fault tolerance can be theoretically achieved via quantum error correction (QEC) \cite{Shor1995,Calderbank1996,Steane1996,Kitaev2003,Terhal2015}, its practical implementation comes with a large hardware overhead \cite{Devitt2013,Fowler2012,OGorman2017} that is out of reach for near-term quantum processors. For the time being, a temporary but precious alternative to QEC is represented by quantum error mitigation (QEM), a collection of techniques which can reduce the impact of errors at the cost of performing extra measurements and post-processing of the data \cite{Temme2017,Li2017,Endo2018b}. Differently from QEC, QEM has been successfully implemented in numerous experiments \cite{Kandala2019,Sagastizabal2019,Google2020,Google2020b} and represents a key resource for the realization of practical applications in the present noisy intermediate-scale quantum (NISQ) era \cite{Preskill2018}. 

To date, a wide spectrum of QEM schemes have been proposed, each one with its own advantages and trade-offs. Importantly, different techniques can be efficiently combined in order to achieve an even better performance \cite{Cai2021,Cai2021practical,Bultrini2021}. Error mitigation schemes include probabilistic error cancellation \cite{Temme2017,Endo2018b}, subspace expansion \cite{McClean2020}, symmetry verification \cite{McArdle2019,Bonet-Monroig2018,Cai2021symm}, purification \cite{Koczor2021,Huggins2021} and zero-noise extrapolation \cite{Li2017,Temme2017,Endo2018b}. The latter is one of the most popular approaches and relies on the possibility to tune the noise level $\lambda$ of the quantum circuit above the smallest achievable value $\lambda_0$. By repeatedly running the circuit at different boosted noise strengths $\lambda_i \geq \lambda_0$, we can thus sample the noisy expectation value of a given observable as a function of $\lambda$ and then try to estimate its zero-noise limit. This operation can be done either by fitting the collected data points with a specific function, typically a single- or multi-exponential \cite{Endo2018b,Cai2021}, or by performing a Richardson-like extrapolation \cite{Richardson,Temme2017}. The second approach has the advantage that it does not require specific assumptions on the $\lambda$-dependence of the expectation values and can be used on a variety of noise models. 
Its effectiveness, however, depends non-trivially on the choice of the different boosted noise strengths $\lambda_i$ \cite{Runge, Hoel1964}.  

In this article, we perform a careful and in-depth analysis of the Richardson method for QEM, highlighting its most important parameters and propose an optimized implementation. With previous methods, the set of noise levels has been determined by a trial and error procedure that terminates once an acceptably small statistical uncertainty is reached. Since for a fixed equidistant spacing of the noise levels, the variance grows exponentially with their number \cite{Giurgica2020}, this was only feasible for no more than 4 noise levels, strongly limiting the mitigation capabilities of Richardson extrapolation. We resolve this issue by demonstrating that the statistical uncertainty can in fact be precisely controlled independently of the number of noise levels. This allows us to significantly improving the mitigation performance, especially for a large sampling budget. Moreover, we find a specific spacing of the noise levels $\lambda_i$ (inspired by the Chebyshev nodes introduced in Ref.\ \cite{Chebyshev1859}) that minimizes, on average, the error on the zero-noise estimation. These discoveries nicely complement the current literature, which mainly focuses on the linear spacing of the $\lambda_i$ and a small number of noise levels \cite{Temme2017,Cai2021practical}. We numerically verify our predictions on different error models, ranging from purely Markovian noise to non-Markovian noise, which can play a significant role in present-day quantum devices \cite{Goswami2021,Morris2019,White2020}.

The article is organized as follows. In Section \ref{sec:Richardson}, we briefly review the Richardson approach to QEM. In Section \ref{sec:findings}, we carefully discuss the three main parameters which characterize the extrapolation scheme and propose our optimized protocol, which we numerically validate in Section \ref{sec:Experiments}. We draw conclusions in Section \ref{sec:Conclusion} and move some technical details to Appendices.

\section{Richardson Extrapolation \label{sec:Richardson}}
Assume that $\rho_\lambda$ is the output state of an imperfect QC, where $\lambda$ is the dimensionless noise parameter that indicates the level of noise in the QC. For this state the expectation value of an observable $A$ is given by 
\begin{align}
    E_\lambda = \mathrm{tr}(\rho_\lambda A).
\end{align}
Since the QC is imperfect, the noise level $\lambda$ cannot be decreased below some finite value $\lambda_0 > 0$ and the limit 
\begin{align}
    \lim_{\lambda \to 0} E_\lambda = E^*
\end{align}
cannot be accessed experimentally. 
However, the noise parameter $\lambda$ can be artificially increased, e.g., by rescaling coupling parameters \cite{Temme2017}, inserting identities into the circuit that are subject to the same noise \cite{Dumitrescu2018, Giurgica2020, He2020, LaRose2021} or gate Trotterization \cite{Schultz2022}. Finding the best way to change the noise levels is an active research area \cite{Schultz2022} and depends on the precise implementation of the quantum computer.  In the following, we assume that we can arbitrarily choose the noise parameter $\lambda = \lambda_0 x$ where $x \in [1, \infty)$. 

The idea behind Richardson extrapolation is to sample $E_{\lambda_0}(x) := E_\lambda$ at $n+1$ distinct \textit{nodes} $x_0, \dots, x_n \in [1, \infty)$ and to find the unique $n$-polynomial that fits these samples. This polynomial approximation of $E_{\lambda_0}(x)$ can be evaluated at $x=0$ which gives an estimate $R_n$ of the noiseless expectation value $E^*$.  For simplicity, we assume that $1 = x_0 < x_1 < \dots < x_n $.

This result of the extrapolation $R_n$ can be directly calculated as a linear combination of the samples $E_{\lambda_0}(x_j)$
\begin{align}
    R_n = \sum_{j=0}^n E_{\lambda_0}(x_j) \gamma_j,
\end{align}
where the factors 
\begin{align}
    \gamma_j = \prod_{k\neq j} \frac{x_k}{x_k-x_j} 
\end{align}
are the Lagrange polynomials at $x=0$. They have the property that $\sum_{j=0}^n\gamma_j = 1$ and $\sum_{j=0}^n \gamma_j x_j^k = 0$ for $k=1,\dots,n$. If we write $E_{\lambda_0}(x)$ as
\begin{align}
 	E_{\lambda_0}(x) = E^* + \sum_{k=1}^\infty a_k \lambda_0^k x^k
\end{align}
for some constant coefficients $a_k$, then the mitigated result reads
\begin{align}
 	R_n  = E^* + \sum_{k=n+1}^\infty a_k \lambda_0^k \sum_{j=0}^n \gamma_j x_j^k.
\end{align}
Compared to the result without mitigation $R_0 = E_{\lambda_0}(1)$, which corresponds to $n=0$, Richardson extrapolation eliminates $a_1, \dots, a_n$ as well as the first $n$ orders of $\lambda_0$.
Therefore, if $\lambda_0$ is sufficiently small then Richardson extrapolation is a promising candidate for QEM.

A general formula for the bias $\mathrm{Bias}[\hat{R}_n] = R_n - E^*$ of this estimate comes from the theory of polynomial approximation. If $E_{\lambda_0}(x)$ is $n+1$ times continuously differentiable on $[0, x_n]$ then there exists a $\xi \in [0, x_n]$ such that
\begin{align}
 	\mathrm{Bias}[\hat{R}_n] = (-1)^{n} E_{\lambda_0}^{(n+1)}(\xi) \frac{C_n}{(n+1)!}. \label{eq:bias0}
\end{align}
$E_{\lambda_0}^{(n+1)}(x)$ denotes the $(n+1)$th derivative of $E_{\lambda_0}(x)$ and $C_n = \prod_{j=0}^n x_j$ (see, e.g.,  Refs.\ \cite{Davis, Lagrange, Chebyshev1859}). 

Besides the error due to noise, additional sampling errors must be considered whenever a quantum mechanical measurement is conducted. If for the noise level $\lambda_0 x_j$ the observable $A$ is measured $N_j$ times, then the sample mean $\hat{E}_{\lambda_0}(x_j)$ has a variance of
\begin{align}
    \mathrm{Var}[\hat{E}_{\lambda_0}(x_j)] = \frac{\sigma_j^2}{N_j} \label{eq:singleVariance}
\end{align}
where $\sigma_j = \sqrt{\mathrm{tr}(\rho_{\lambda_0 x_j} A^2) - E_{\lambda_0}(x_j)^2}$. The $\sigma_j$'s depend on the observable $A$ as well as on the expected value $E_{\lambda_0} (x_j)$ and the density matrix $\rho_{\lambda_0 x_j}$ of the quantum state. They are unknown quantities before the experiment has been conducted, but it is reasonable to assume the $\sigma_j$ to be of the same order of magnitude. For simplicity, we assume that $\sigma_j = \sigma =$ constant for all $j$ in the following. 

Since the result of Richardson extrapolation $R_n$ is a linear combination of all $E_{\lambda_0}(x_j)$, the corresponding estimator $\hat{R}_n$ is thus also affected by sampling errors
\begin{align}
    \mathrm{Var}[\hat{R}_n] &= \sum_{j=0}^n \gamma_j^2 \frac{\sigma^2}{N_j} \label{eq:variance0}.
\end{align}
As we will see later, this leads to an increase in variance compared to the unmitigated result with the same number of measurements $\mathrm{Var}[\hat{R}_0] = \frac{\sigma^2}{\sum_{j=0}^n N_j}$. 

\begin{figure}
    \centering
    \includegraphics[width=0.475\textwidth]{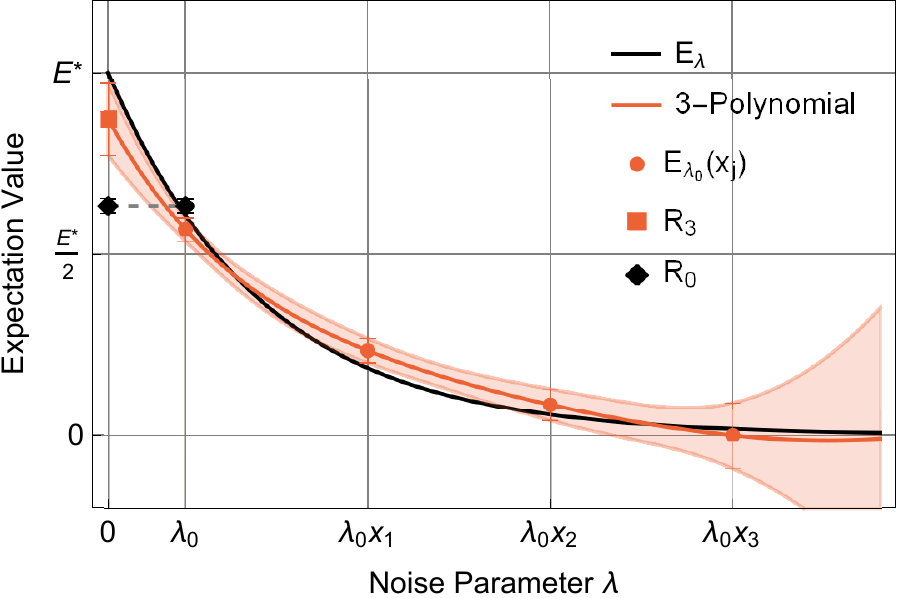}
    \caption{Example for Richardson extrapolation with $n=3$ for a noisy expectation value which decays exponentially $E_\lambda = E^* e^{-\lambda}$. The black diamond is the result without mitigation $R_0$ and the red square the mitigated result $R_3$, both obtained with the same sampling budget. The colored interval indicates the standard deviation of the extrapolation using Lagrange polynomials. The error bars are the standard deviations of the samples $E_{\lambda_0}(x_j)$ adjusted according to Eqs.\ \eqref{eq:singleVariance} and \eqref{eq:Polynomial_Approx}. The nodes lie equidistant.
    \label{fig:Polynomial_Approx}}
\end{figure}

In conclusion, Richardson extrapolation is able to reduce the bias of the estimate of $E^*$, however, only at the cost of increasing its variance. An example that illustrates these effects is shown in Fig.\ \ref{fig:Polynomial_Approx}.

If we choose a linear spacing of the nodes 
\begin{align}
	x_j = 1 + j, \label{eq:dumblinear}
\end{align}
then $\mathrm{Var}[\hat{R}_n]$ grows exponentially with $n$ in the limit $n \to \infty$ \cite{Giurgica2020}. This poses the following questions:
\begin{enumerate}
 \item Is there a way to reduce this increase in $\mathrm{Var}[\hat{R}_n]$?
 \item Is there a scaling method for $x_j$ that optimizes the tradeoff between bias and variance?
 \item Is there an optimal $n$?
\end{enumerate}
In the following section, we investigate these questions and how to efficiently execute this mitigation technique, i.e., how to choose the free parameters $n,\{N_j \}_{j=0\dots,n}$ and $\{ x_j \}_{j=0\dots,n}$. For a fixed variance we search for those parameters that, given the same computational resources, minimize the bias of the estimator $\hat{R}_n$.

\section{Optimized QEM Protocol \label{sec:findings}}

In the literature, it has been noted that the numbers of measurements $N_j$ for a given noise level $\lambda_0 x_j$ should be adjusted according to their weights $\gamma_j$ in Eq.\ \eqref{eq:variance0} \cite{Hoel1964, Cai2021practical}. This is also the starting point of our analysis in Section \ref{sec:N_j}. After that, in Section \ref{sec:nodes}, we turn to the spacing of the nodes $x_j$. Different strategies have been considered before, including equidistant and exponential nodes as well as nodes based on the extrema of the Chebyshev polynomials of the first kind. We compare these spacing methods to a new set of nodes which we show to be the optimal ansatz for QEM. In Section \ref{sec:NrNodes}, we then focus on the number of nodes $n$. We show that, for certain noise functions and a proper choice of the nodes, increasing $n$ can substantially improve mitigation.
The complete procedure is summarized in Section \ref{sec:summary}.

\subsection{Number of measurements \texorpdfstring{$N_j$}{TEXT} and measurement overhead \texorpdfstring{$\Lambda$}{TEXT} \label{sec:N_j}}

A first simple step for the optimization of Richardson extrapolation is to reduce $\mathrm{Var}[\hat{R}_n]$ by adjusting $N_j$ while keeping the total sampling budget $\sum_{j=0}^n N_j= N_\mathrm{tot}$ constant \cite{Hoel1964, Cai2021practical}. Whenever non-integer values for $N_j$ appear in the following they should be rounded to the next nearest integer. A simple calculation shows that $\mathrm{Var}[\hat{R}_n]$ is minimized by
\begin{align}
 	N_j = N_\mathrm{tot} \frac{|\gamma_j|}{\sum_{k=0}^n|\gamma_k|}. \label{eq:Polynomial_Approx}
\end{align}
Defining $\Lambda = \sum_{j=0}^n |\gamma_j|$ and $N_\mathrm{eff} = N_\mathrm{tot}/\Lambda^2$ the minimized variance of $\hat{R}_n$ can be written as
\begin{align}
 	 \mathrm{Var}[\hat{R}_n]=\frac{\sigma^2}{N_\mathrm{tot}} \Lambda^2 = \frac{\sigma^2}{N_\mathrm{eff}}. \label{eq:variance1}
\end{align}
This way the estimator $\hat{R}_n$ that uses $N_\mathrm{tot}$ measurements has the same variance as the sample mean of a single expectation value that is measured $N_\mathrm{eff}$ times. 
$\Lambda^2 \geq 1$ is therefore the minimal overhead in measurements required to compensate for Richardson extrapolation.
An example for this procedure is given in Fig.\ \ref{fig:Polynomial_Approx}, where the standard deviation of each sample $E_{\lambda_0}(x_j)$ is adjusted according to Eq.\ \eqref{eq:Polynomial_Approx}. In this case, the nodes are chosen such that $\Lambda = 5$.

The quantities $N_\mathrm{tot}$ and $N_\mathrm{eff}$ can be chosen freely at the beginning of an experiment. The first one is the sampling budget of the experiment while the second determines the variance of the estimator $\hat{R}_n$. The sampling overhead is thus given by $\Lambda^2 = \frac{N_\mathrm{tot}}{N_\mathrm{eff}}$. 
As we will see later, this overhead has a major influence on the range of noise that can be mitigated. After that, the nodes $\{x_j\}_{j=0,\dots, n}$ must be chosen in such a way that $\sum_{j=0}^n |\gamma_j| = \Lambda$. For each spacing method, the nodes usually only depend on $x_1$  (compare, e.g., Eqs.\ \eqref{eq:nodesLin} - \eqref{eq:nodesTC}). It is therefore possible to first choose a spacing method and then numerically solve this equation to get the value of $x_1$ that leads to the chosen $\Lambda$. Importantly, this can be done independently of $n$. The variance $\mathrm{Var}[\hat{R}_n]$ is therefore independent of the number of nodes $n$ and it is not necessary to restrict considerations to small $n$.

Typically, a total sampling budget of the order $10^5$ to $10^{6}$ \cite{Kandala2019, Mari2021, Czarnik2021} or even up to $10^{10}$ measurements \cite{Bultrini2021} is considered in the literature. If, for example, we have a budget of $N_\mathrm{tot} = 10^6$ measurements and choose $N_\mathrm{eff} = 1024$ then the sampling overhead is about $\Lambda^2 = 32^2$ . 

Now, we can compare different spacing methods in a \textit{fair} way, i.e., for the same sampling budget $N_\mathrm{tot}$ as well as the same variance $\sigma^2/N_\mathrm{eff}$ of the estimator $\hat{R}_n$ of $E^*$. This way, we can search for those nodes that minimize the bias of $\hat{R}_n$.

\subsection{Spacing of the nodes \texorpdfstring{$x_j$}{TEXT} \label{sec:nodes}}
Taking a closer look at the bias in Eq.\ \eqref{eq:bias0} and the variance in Eq.\ \eqref{eq:variance1}, we observe that the quantities that depend on the spacing of the nodes are 
\begin{align}
 	C_n &= \prod_{j=0}^n x_j
 	\intertext{and}
 	\Lambda &= \sum_{j=0}^n |\gamma_j| = \sum_{j=0}^n |\prod_{k\neq j} \frac{x_k}{x_k-x_j}|.
\end{align}
If all $x_j$ are small, then $C_n$ and the bias are small, however, at the same time, all $x_j$ lie close together such that $\Lambda$ and the variance get large. On the contrary, if all $x_j$ are far apart, then $\Lambda$ is close to 1 but $C_n$ gets large. 
The aim of this section is to determine nodes that find the best compromise between these effects. 

In the literature the equidistant (linear - L), exponential (E) and Chebyshev nodes are frequently used. The latter requires a word of caution, as the expression \textit{Chebyshev nodes} or \textit{points} has been used both for the zeros and the extrema of the Chebyshev polynomials of the first kind. While the zeros are well suited for \textit{interpolation} \cite{Chebyshev1859, Brutman1978}, we are more interested in \textit{extrapolation} and therefore focus on the extrema (C). The different spacing methods read:
\begin{align}
    x_j^L &= 1 + j(x_1 -1) \label{eq:nodesLin} \\
    x_j^E &= x_1^j \label{eq:nodesExp} \\
    x_j^{C} &= 1+ \frac{\sin^2(\frac{j}{n} \frac{\pi}{2})}{\sin^2(\frac{1}{n} \frac{\pi}{2})}(x_1-1) \label{eq:nodesCheb2}
\end{align}
where $j=0,\dots,n$.

Note, that, in contrast to Eq.\ \eqref{eq:dumblinear}, Eq.\ \eqref{eq:nodesLin} has a variable distance between two neighboring nodes. By choosing $N_\mathrm{eff}$ and adjusting this distance to $\Lambda^2 = \frac{N_\mathrm{tot}}{N_\mathrm{eff}}$ as discussed in Sec.\ \ref{sec:N_j}, the problem of variances increasing exponentially with $n$ is resolved. 

From Ref.\ \cite{Hoel1964}, we know that $x_j^{C}$ minimize $x_n$ for a given $\Lambda$. However, they do not at the same time minimize the product $C_n$ small which is why we look at the following optimization problem:

For constant $\Lambda = \sum_{j=0}^n|\gamma_j|$ and $x_0=1$ find the nodes $x_0, \dots, x_n$ that minimize $C_n$. 
This optimization problem is solved by the nodes
\begin{align}
    x_j^T = 1 + \frac{\sin^2(\frac{j}{n+1} \frac{\pi}{2})}{\sin^2(\frac{1}{n+1} \frac{\pi}{2})}(x_1-1). \label{eq:nodesTC}
\end{align}
A proof is outlined in Appendix \ref{sec:Appendix}.

\begin{figure}
    \centering
    \includegraphics[width=0.45\textwidth]{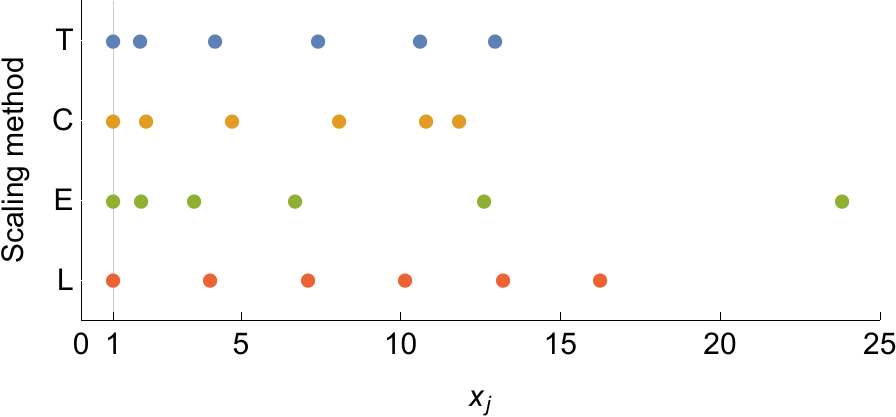}
    \caption{Comparison of the different spacing methods for $n=5$ and $\Lambda = 10$. The scaling methods are: tilted Chebyshev (T), extremal Chebyshev (C), exponential (E) and equidistant (L). \label{fig:nodes_comparison}}
\end{figure}

This result resembles the (extremal) Chebyshev nodes where $n$ is increased by 1 and the last node is left out. This asymmetry makes sense since the goal is to extrapolate \textit{to the left} and precision \textit{to the right} is subordinate. We call the nodes \textit{tilted Chebyshev nodes} (T). Fig.\ \ref{fig:nodes_comparison} provides a visual comparison between the different spacing methods.

\begin{figure}
    \centering
    \includegraphics[width=0.55\textwidth]{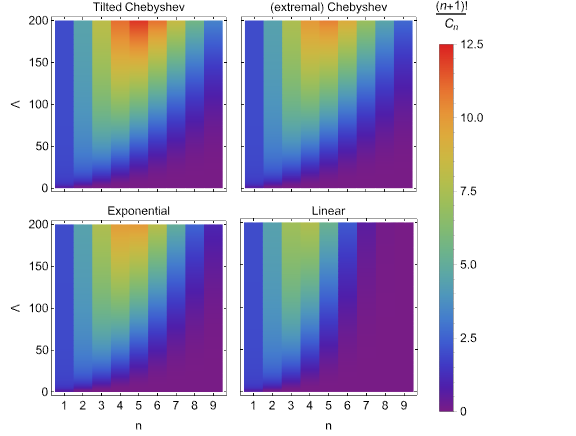}
    \caption{The ratio $\frac{(n+1)!}{C_n}$ for different values of $\Lambda$ and $n$ for the spacing methods $x_j^T$, $x_j^{C}$, $x_j^E$ and $x_j^L$. \label{fig:Density_Grid}}
\end{figure}

As a numerical verification of this result, we plot $\frac{(n+1)!}{C_n}$ for different spacing methods and numbers of nodes over $\Lambda$ in Fig.\ \ref{fig:Density_Grid}. The larger this ratio, the smaller is the bias of $\hat{R}_n$ and the better the result of Richardson extrapolation. From this plot, we observe that the tilted Chebyshev nodes indeed lead to a lower $C_n$ and consequently a higher ratio $\frac{(n+1)!}{C_n}$ than the other methods. While for $n \leq 3$ all scaling methods perform similarly, the tilted Chebyshev nodes perform best for larger $n$. For example for $n=7$ they lead to a $C_n$ that is about 1.25, 2 and 35 times smaller than that of the (extremal) Chebyshev, exponential and linear nodes, respectively. Remember, that this result is independent of the noise model and the improvement does not require any additional resources.

Since we have determined $N_j$ and $x_j$, the last remaining parameter of Richardson extrapolation is the number of nodes $n$.

\subsection{Number of nodes \texorpdfstring{$n$}{TEXT} \label{sec:NrNodes}}
While the variance $\mathrm{Var}[\hat{R}_n]$ is independent of the number of nodes, the bias $\mathrm{Bias}[\hat{R}_n]$ is not. It depends on $E_{\lambda_0}^{(n+1)}(\xi)$ as well as on $\frac{C_n}{(n+1)!}$, see Eq.\ \eqref{eq:bias0}. Without knowing the whole function $E_{\lambda_0}(x)$, we cannot predict the derivative $E_{\lambda_0}^{(n+1)}(\xi)$ and thus cannot identify the best value for $n$. 

Useful insight for the choice of $n$ can be gained with the (strong) assumption that all $E^{(n+1)}_{\lambda_0}(\xi)$ are approximately the same for all $n$. In this case, the $n$ dependence of the bias is completely determined by $\frac{C_n}{(n+1)!}$. For a given spacing method, this quantity can be analyzed as a function of $n$ and $\Lambda$. The results for the discussed spacing methods are shown in Fig.\ \ref{fig:Density_Grid}, where we plot the ratio $\frac{(n+1)!}{C_n}$.

Given the assumption that $|E_{\lambda_0}^{(n+1)}(\xi)|$ is approximately constant, Fig.\ \ref{fig:Density_Grid} could be used to determine the optimal $\hat{n}$ once $\Lambda$ is chosen. 
However, different derivatives have different properties in reality and $\xi$ can be very different for different $n$ such that the meaning of this \textit{optimal} $\hat{n}$ should not be overestimated. For example, if $|E_{\lambda_0}^{(n+1)}(\xi)|$ decreases with $n$ then this favors larger $n \geq \hat{n}$. 
While there is no optimal $n$ in gerneral, Fig.\ \ref{fig:Density_Grid} might be able to suggest an interval from which we choose $n$ in order to minimize the bias.

In practice, there might be additional factors limiting the choice of $n$. For example, if $\Lambda$ is small then $n$ should not be chosen too large, otherwise, $x_n$ can get exceedingly large. Furthermore, this can lead to very low measurement numbers $N_j$ which can be avoided for example by decreasing $n$ or setting a lower bound on $N_j$. On the contrary, if $n$ is too small while $\Lambda$ is large then the $x_j$ lie close to each other which makes the procedure susceptible to errors in the amplification factors $x_j$.

In the next section we summarize the procedure before turning to numerical simulations of examples.

\subsection{Summary of the procedure \label{sec:summary}}
We describe above that the relevant parameters of Richardson extrapolation are $N_\mathrm{eff}$, the overhead $\Lambda^2$ and the number of nodes $n$. Since the spacing method cannot always be chosen at will we also include it into this list. We now summarize our protocol for Richardson extrapolation:
\begin{enumerate}
\item Define a variance $\mathrm{Var}[\hat{R}_n] = \sigma^2/N_\mathrm{eff}$ that is acceptable and choose $N_\mathrm{eff}$ accordingly.
\item Choose  $\Lambda^2$ depending on the sampling budget ($N_\mathrm{tot} = N_\mathrm{eff}\Lambda^2$). The larger, the better\footnote{There is one exception to this rule. If errors on the nodes $x_j$ play a significant role, it can make sense to choose a smaller $\Lambda$ to reduce the effect of the errors on $x_j$ for mitigation.}.
\item Determine the spacing of the nodes. As discussed in Section \ref{sec:nodes}, the tilted Chebyshev nodes most likely lead to the best extrapolation result. 
\item Pick a value for $n$. Given $\Lambda$ and the spacing method, $\hat{n}$ is the $n$ maximizing $\frac{(n+1)!}{C_n}$ in Fig.\ \ref{fig:Density_Grid} and can be viewed as guide for a good choice of $n$. If $E_{\lambda_0}(x)$ and its derivatives vary a lot, then $n \approx \hat{n}$ should be a good choice whereas if it has a small slope or is an exponential decay, $n$ should be chosen larger $n \geq \hat{n}$. 
\end{enumerate}

\begin{figure}
	\centering
	\includegraphics[width=0.5\textwidth]{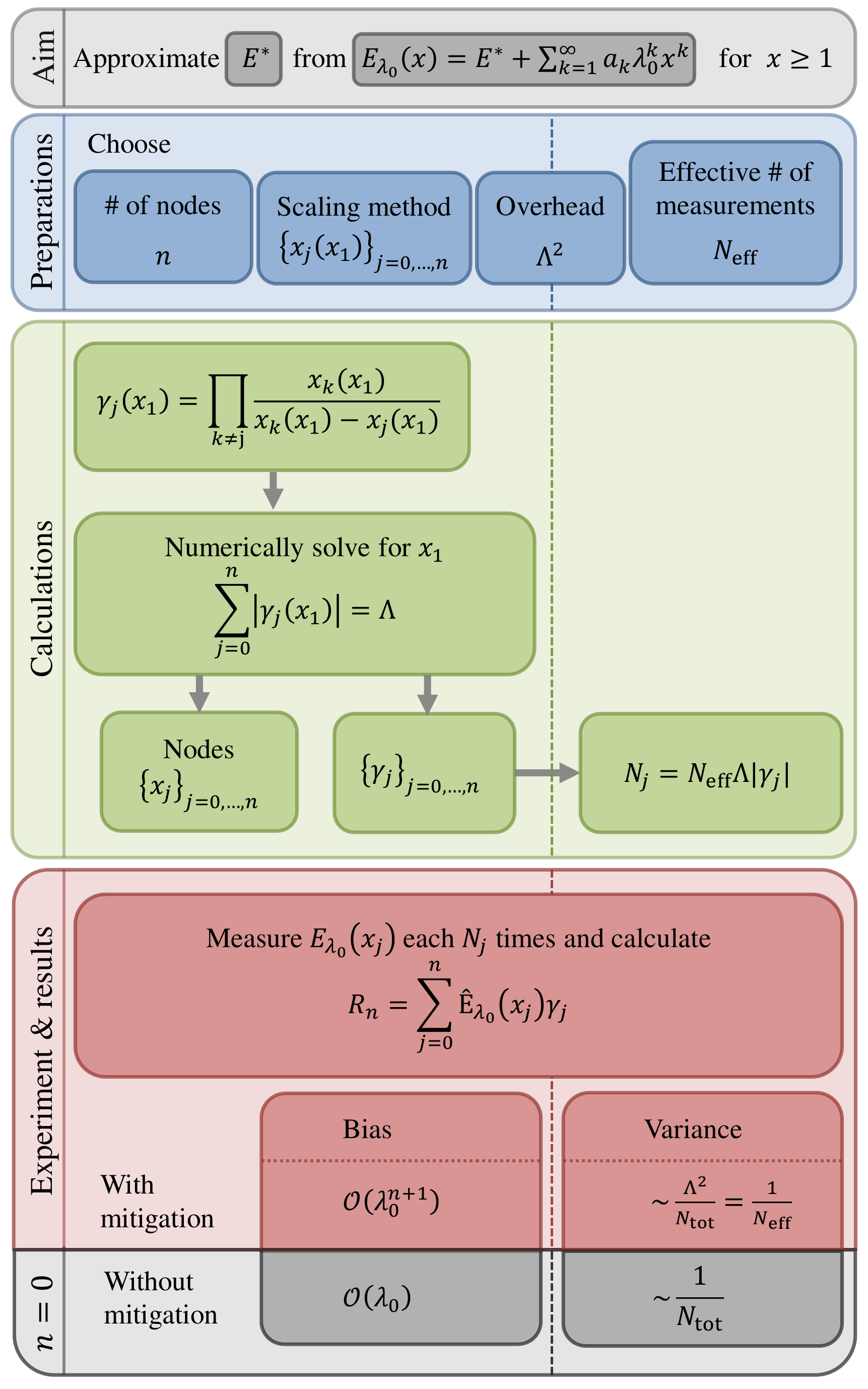}
	\caption{Schematic representation of the procedure for Richardson extrapolation. \label{fig:ZNEScheme}}
\end{figure}

Once these choices have been made, the procedure is straightforward. First, numerically solve $\Lambda = \sum_{j=0}^n | \gamma_j|$ for $x_1$. Use this to calculate the nodes $x_j$, $\gamma_j$, and $N_j$. Measure $E_{\lambda_0}(x_j)$ each $N_j$ times and calculate the estimate $R_n = \sum_{j=0}^n \gamma_j E_{\lambda_0}(x_j)$ of $E^*$. This estimate has a statistical error (standard deviation) of $\sigma / \sqrt{N_\mathrm{eff}}$. This procedure is schematically represented in Fig.\ \ref{fig:ZNEScheme}.

\section{Examples \label{sec:Experiments}}
In this section, we test the procedure for Richardson extrapolation on different noise models and investigate the dependency on $n$. First, we focus on Markovian noise and in a second step address non-Markovian noise. All estimators in this section have a variance of $\sigma^2/N_\mathrm{eff}$. Note that the following figures focus on the bias and do not show the variance. 

Markovian noise (M) can emerge for example when the quantum state interacts with a bath and the bath correlation functions decay much faster than typical times scales of the system \cite{Breuer}. It typically leads to an exponential decay of the expectation value \cite{Cai2021practical}
\begin{align}
    E_{\lambda_0}^\mathrm{M}(x) = E^* e^{-\lambda_0 x}. \label{eq:Markov}
\end{align}
\begin{figure}
	\centering
	\includegraphics[width=0.48\textwidth]{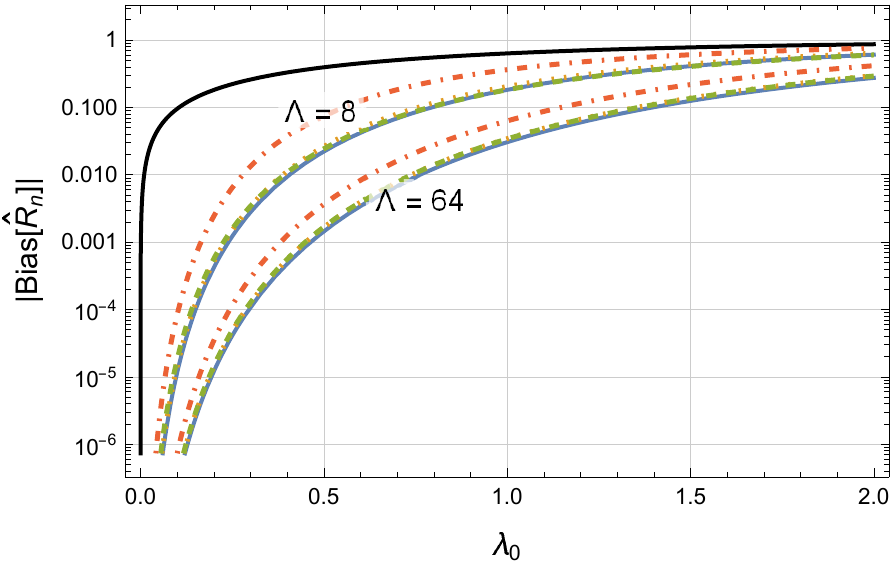}
	\caption{Richardson extrapolation used on the Markovian noise model in Eq.\ \eqref{eq:Markov}. The bias is shown over the minimal noise strength $\lambda_0$ for $\Lambda = 8$ and 64 and $n=5$. The spacing methods are: tilted Chebyshev (blue, solid), extremal Chebyshev (yellow, dotted), exponential (green, dashed) and equidistant (red, dot-dashed). The black solid line shows the result without mitigation $|\mathrm{Bias}[\hat{R}_0]|$.  \label{fig:Markov_lambda}}
\end{figure}
For convenience we assume it to be dimensionless and choose $E^*=1$. For $\Lambda=8$ and $64$ and $n=5$, $|\mathrm{Bias}[\hat{R}_n]|$ is plotted over $\lambda_0$ in Fig.\ \ref{fig:Markov_lambda}. We can see that the bias depends most notably on the overhead $\Lambda^2$. The larger $\Lambda^2$ the 
It can therefore make sense to use more measurements and increase $\Lambda^2$ in order to reduce the bias. 
Furthermore, we can see that for $n=5$ the equidistant spacing performs significantly worse than the other methods in accordance with the general results presented in Section \ref{sec:findings}.

\begin{figure}
	\centering
	\includegraphics[width=0.48\textwidth]{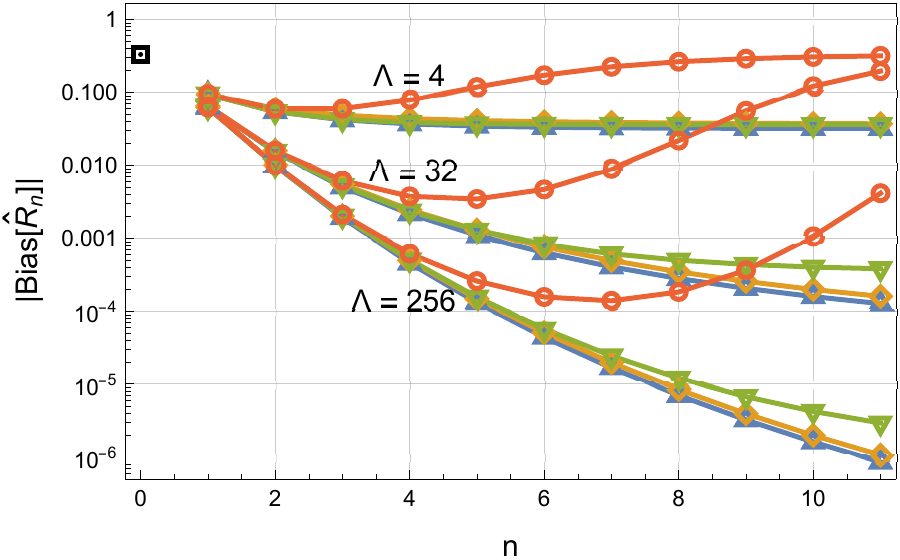}
	\caption{Comparison of the bias of Richardson extrapolation on Markovian noise for fixed $\lambda_0 = 0.4$ and different $n$. The overhead in measurements is $\Lambda^2 = 4^2, 32^2$ and $256^2$, respectively. Note that the smaller the bias the better the QEM protocol works. The different spacing methods are: tilted Chebyshev (blue, uptriangle), extremal Chebyshev (yellow, diamond), exponential (green, downtriangle) and equidistant (red, circle). The black square for $n=0$ is the bias without mitigation $|\mathrm{Bias}[\hat{R}_0]|$. \label{fig:Markov_n}}
\end{figure}

In order to investigate the dependency on $n$, we fix $\lambda_0 = 0.4$ and plot the resulting estimate for different $n$ in Fig.\ \ref{fig:Markov_n}. Remember that, for a given choice of $\Lambda$, each data point is calculated using the same sampling budget $N_\mathrm{tot}$, independently of $n$. We can observe that the bias can be significantly reduced by increasing $n$ for all spacing methods except for the equidistant nodes. 

For this noise model, increasing $n$ reduces the error by more than an order in magnitude. 
Interestingly, this is the case even if $\lambda_0 \geq 1$, 
(which is not shown here) 
such that this effect is not solely caused by a decrease of $|E^{(n+1)}_{\lambda_0}(\xi)|$ with $n$. It seems that increasing $n$ is advantageous for any noise strength if the noise is given by an exponential decay. 
In practice, the limitations mentioned in Section \ref{sec:NrNodes} render $n \gg 10$ impractical for many QEM applications.

If we know the explicit form of the noise function, we can directly fit it to the data points and there is no need for Richardson extrapolation. In the case that the noise is purely Markovian and can be approximated by a single or multi-exponential decay, this has been done e.g., in Refs.\ \cite{Endo2018b, Cai2021}. Another idea is to use Richardson extrapolation on the data points $\log E_{\lambda_0}(x_j)$ for an estimate of $\log E^*$ (assuming $\lim_{\lambda \to \infty} E_\lambda = 0$) which is equivalent to a poly-exponential extrapolation \cite{Giurgica2020}. However, if the noise is a combination of different functions or is unknown altogether, then an important advantage of Richardson extrapolation is that polynomials are flexible enough to fit different kinds of functions. 

To test this we now introduce non-Markovian (NM) noise components to the noise function. Therefore, we use the following model
\begin{align}
    E_{\lambda_0}^{\mathrm{NM}}(x)= e^{-(1-\eta) \lambda_0 x} \Big(&\cos( \eta\lambda_0 x)\cos(\omega) \notag \\ 
    + \frac{\eta\lambda_0 x}{\omega} &\sin( \eta\lambda_0 x)\sin(\omega) \Big) \label{eq:nonMarkov}
\end{align}
where $\omega = \sqrt{4 + (\eta\lambda_0 x)^2}$. The parameter $\eta$ can be used to interpolate between purely Markovian noise ($\eta = 0$) and highly non-Markovian noise ($\eta = 1$).
A more detailed description of this noise model can be found in Appendix \ref{sec:AppendixNM}. Notice that, for this specific noise model, we have $E^* = \cos(2) \approx -0.416$.

\begin{figure}
	\centering
	\includegraphics[width=0.48\textwidth]{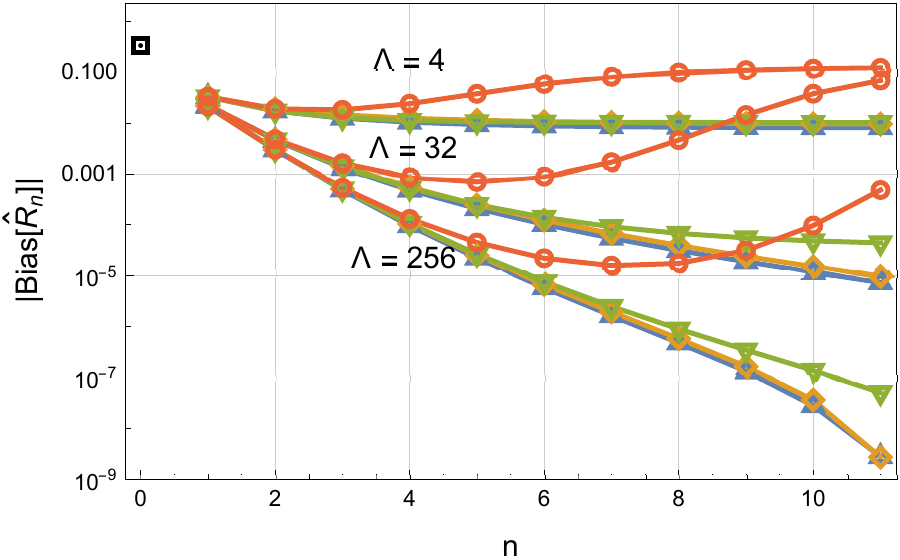}
	\caption{Comparison of the bias of Richardson extrapolation on the noise model in Eq.\ \eqref{eq:nonMarkov} with $\eta = 0.1$ and $\lambda_0 = 0.4$ for different $n$. The overhead in measurements is $\Lambda^2 = 4^2, 32^2$ and $256^2$, respectively. Note that the smaller the bias the better the QEM protocol works. The different spacing methods are: tilted Chebyshev (blue, uptriangle), extremal Chebyshev (yellow, diamond), exponential (green, downtriangle) and equidistant (red, circle). The black square for $n=0$ is the bias without mitigation $|\mathrm{Bias}[\hat{R}_0]|$. \label{fig:NonMarkov_n02}}
\end{figure}

In Fig.\ \ref{fig:NonMarkov_n02}, we consider the presence of a small non-Markovian component, $\eta = 0.1$, and plot the bias of $\hat{R}_n$ over $n$.
In this case, the error mitigation behaves similarly to the purely Markovian case in Fig.\ \ref{fig:Markov_n}. In particular, for all spacing methods except the equidistant nodes, the bias can be reduced significantly by increasing $n$. 
For these simple Markovian or near-Markovian noise models it seems that there is no disadvantage when increasing $n$.  

\begin{figure}
	\centering
	\includegraphics[width=0.48\textwidth]{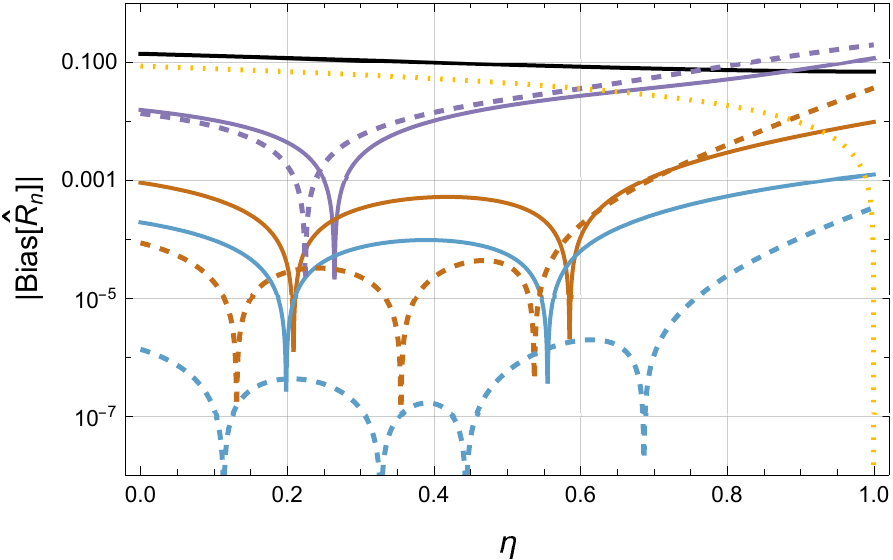}
	\caption{The bias of mitigation for different levels of non-Markovianity. The noise is purely Markovian for $\eta = 0$ and highly non-Markovian for $\eta=1$. There are three different values for $\Lambda$: 4 (purple), 32 (brown) and 256 (blue) and two for $n$: 4 (solid) and 9 (dashed).  The black solid line is the result without mitigation $|\mathrm{Bias}[\hat{R}_0]|$. The dotted yellow line is the fake-node approach with $S(x)=x^2$ for $\Lambda = 4$ and $n=9$. For some $\eta$, there is a transition between over- and underestimation of the true expectation value $E^*$ which is why the bias can become 0.
	\label{fig:MarkovNonMarkov49}}
\end{figure}

In Fig.\ \ref{fig:MarkovNonMarkov49}, we focus on the tilted Chebyshev nodes and investigate how mitigation for different $n$ changes when interpolating between Markovian and non-Markovian noise. For $\eta \approx 0$, $n=9$ (dashed lines) clearly leads to a lower bias than $n=4$ (solid lines). For $\eta \to 1$ this is not true anymore, especially for small $\Lambda$. For the highly non-Markovian noise model, the derivatives $|E_{\lambda_0}^{(n+1)}(\xi)|$ do not decrease (on average) when increasing $n$. From the considerations in Section \ref{sec:NrNodes} we therefore expect the mitigation result to worsen when $n \gg \hat{n}$ which is what can be observed in the plot. This confirms, that for a simple noise model, $n$ should be chosen larger than $\hat{n}$ whereas for more complex noise $n$ should be chosen in proximity to $\hat{n}$. The full dependency of the bias on $n$ for the highly non-Markovian case $\eta = 0.9$ is shown in Appendix \ref{sec:AppendixB}.

Finally, we want to briefly comment on the idea of \textit{fake nodes} introduced in Ref.\ \cite{FakeNodes}. From the Runge problem \cite{Runge}, it is well known that equidistant nodes are not well suited for interpolation. 
However, if specific hardware requirements force the nodes to be equidistant, their limitations can be bypassed by mapping them to so called \textit{fake nodes} $\tilde{x}_j = S(x_j)$ for some invertible $S(x)$ with $S(0)=0$ and $S(1) = 1$ and performing Richardson extrapolation of $E_{\lambda_0}(S^{-1}(\tilde{x}))$ on these fake nodes. By choosing for examples the zeros of the Chebyshev polynomials as fake nodes the limitations of equidistant nodes can be avoided. While this proved to be effective for the interpolation of the Runge function, the map to the tilted Chebyshev nodes of the same $\Lambda$ does not provide a benefit for the Richardson extrapolation of Eqs. \eqref{eq:Markov} or \eqref{eq:nonMarkov}. However, the same formalism can be used in a different way to improve the mitigation result in the completely non-Markovian case $\eta = 1$. Since then the function $E_{\lambda_0}^{NM}(x)$ is an even function in $x$, it makes sense to use only even polynomials for extrapolation. This idea has been mentioned in Ref.\ \cite{Temme2017} and can be achieved by choosing $S(x) = x^2$, for which $E_{\lambda_0}^{NM}(x)$ is approximated in the space spanned by $\{1, x^2, x^4, \dots, x^{2n} \}$ instead of $\{1, x, x^2, \dots, x^{n} \}$. An example for this with $n=9$ is plotted in Fig.\ \ref{fig:MarkovNonMarkov49}. There the fake nodes $\tilde{x}_j$ are the tilted Chebyshev nodes with $\Lambda=4$ and the corresponding real nodes are $x_j = \sqrt{\tilde{x}_j}$.

\section{Conclusion \label{sec:Conclusion}}
The goal of quantum error mitigation is to find an estimator of a noiseless expectation value from noisy observations.  This estimator should have small bias and variance for as little computational resources as possible. We analyze the interplay of these three factors in the case of Richardson extrapolation. It is a common perception that Richardson extrapolation is limited by its substantial increase in variance, especially if the number of nodes is increased. We propose a protocol in which the variance is independent of the number of nodes and can be tuned precisely. This allows to increase the number of nodes $n$, which in turn can reduce the bias of the zero-noise estimate significantly, especially for large sampling overheads. Furthermore, we show that equidistant nodes are not a good choice for Richardson extrapolation and propose the tilted Chebyshev nodes which optimize the trade-off between bias and variance and outperforms other choices, especially for large $n$.

\section{Acknowledgments}
This work was supported by the Würzburg-Dresden Cluster of Excellence on Complexity and Topology in Quantum Matter (EXC2147, project ID 390858490) and by the DFG (SPP1666 and SFB1170 \ldq ToCoTronics\rdq). We also acknowledge financial support by the Bavarian High Tech Agenda.

\bibliographystyle{unsrtnat}
\bibliography{Ref}

\begin{appendix}
\section{Tilted Chebyshev nodes \label{sec:Appendix}} 
The optimization problem described in Section \ref{sec:nodes} is formalized as
\begin{align}
 	\underset{x_0,\dots,x_n \in [1, \infty)}{\mathrm{minimize}}&~~C_n  \label{eq:optimization_real1}\\
 	\mathrm{subject~to}&~~ \sum_{j=0}^n |\gamma_j| = \Lambda \\
 	\mathrm{and}&~~ x_0~ = 1. \label{eq:optimization_real3}
\end{align}
It can be solved using Lagrange multipliers $\mu$ and $\mu_0$. The solution has to satisfy
\begin{align}
 	\nabla \left( C_n - \mu \left( \sum_{j=0}^n |\gamma_j| - \Lambda \right) - \mu_0 (x_0 - 1) \right) = 0
\end{align}
where $\nabla$ is the gradient with respect to $(x_0, \dots, x_n, \mu, \mu_0)^T$. Performing the derivatives, this leads to
\begin{align}
 	C_n = \mu_0 x_0 \delta_{k,0} + \mu \varphi_k
\end{align}
for $k=0,\dots, n$, where $\delta_{k,0}$ is the Kronecker-Delta and
\begin{align}
 	\varphi_k := \sum_{j\neq k} \frac{1}{x_j - x_k} \left( (-1)^j x_j \gamma_j + (-1)^k x_k \gamma_k \right).
\end{align}
Notice that $\gamma_j = (-1)^j |\gamma_j|$. Summing this expression over $k=0,\dots, n$, this leads to
\begin{align}
 	\sum_{k=0}^n C_n = (n+1)C_n = \mu_0
\end{align}
due to the antisymmetry under exchange of $j \leftrightarrow k$ and thus
 \begin{align}
	\mu \varphi_k = 
	\begin{cases}
		- n C_n ~~~~ &\mathrm{if} ~~ k=0\\
		C_n ~~~~ &\mathrm{else}.
	\end{cases} \label{eq:optimization}
\end{align}
What is left to do is to show that Eq.\ \eqref{eq:nodesTC}, i.e.,
\begin{align}
 	x_j = 1 + \frac{\sin^2(\frac{j}{n+1} \frac{\pi}{2})}{\sin^2(\frac{1}{n+1} \frac{\pi}{2})} (x_1 -1), \label{eq:sin2appendix}
\end{align}
solves this system of equations Eq.\ \eqref{eq:optimization}. In order to do so the following equations are useful
\begin{align}
	\alpha &:= \frac{\pi}{2(n+1)},\\
 	\frac{1}{x_m-x_j} &= \frac{1}{x_1-1} \frac{\sin^2(\alpha)}{\sin^2(m\alpha) - \sin^2(j\alpha)}\\
 	&= \frac{1}{x_1-1} \frac{\sin^2(\alpha)}{\sin((m-j)\alpha)\sin((m+j)\alpha)},\notag\\
 	\frac{\gamma_0}{C_n} &= \left( \prod_{l=0}^n \frac{\sin^2(\alpha)}{\sin^2(l \alpha)}\frac{1}{x_1-1} \right), \\
 	\frac{x_j \gamma_j }{C_n} &= \frac{x_j}{C_n} \prod_{m\neq j} \frac{x_m}{x_m-x_j}\\
 	&= \prod_{m\neq j} \frac{(x_1-1)^{-1} \sin^{2}(\alpha)}{\sin((m-j)\alpha) \sin((m+j)\alpha)}. \notag
\end{align}
In the last product, the sine appears in the denominator $2n$ times. The symmetries $\sin(x)=-\sin(-x)$ as well as $\sin((n+1+k)\alpha) = \sin((n+1-k)\alpha)$ can be used to bring all arguments into the interval $[ 0, \pi/2]$ which leads to an overall sign of $(-1)^j$. Then each argument $l \alpha$ with $l=1,\dots,n$ appears exactly twice except for $l=2j$ and $l=n+1-j$ (only once) and $l=j$ (three times). Additionally, $l=n+1$ appears once. This leads to
\begin{align}
	\frac{x_j \gamma_j }{C_n} &= (-1)^j \frac{\gamma_0}{C_n} \frac{\sin((n+1-j)\alpha) \sin(2j\alpha)}{\sin((n+1)\alpha)\sin(j\alpha)}\\	
	&= (-1)^j \frac{\gamma_0}{C_n} \left( 2 \cos^2(j \alpha) - \delta_{j,0} \right). 
\intertext{such that for arbitrary $j$}
	x_j \gamma_j &= (-1)^j x_0 \gamma_0 \left( 2 \cos^2(j \alpha) - \delta_{j,0} \right).
\end{align}
Inserting all of this into $\varphi_k$, this gives
\begin{align}
 	\varphi_k 
 	&= \sin^2(\alpha) \frac{x_0 \gamma_0}{x_1-1} \Omega_k,
\end{align}
where $\Omega_k := \sum_{j\neq k} \frac{2\cos^2(j \alpha) + 2\cos^2(k \alpha) - \delta_{k,0} - \delta_{j,0}}{\sin^2(j\alpha)-\sin^2(k\alpha)}$. A last step is to show that
\begin{align}
 	\Omega_k =
 	\begin{cases}\!
    		~2n(n+1) ~ &\mathrm{if}~ k=0 \\ 
    		~-2(n+1) ~ &\mathrm{else}.
	\end{cases} \label{eq:fineq}
\end{align}
This last step has not been achieved analytically for arbitrary $n$. Only for $k=0$ a proof can be found for example in Ref.\  \cite{proofeq23} Ch. IX, Art. 69. However, for a given $n$ this formula can be easily checked numerically which we have done for $n \leq 1000$. For the purpose of zero noise extrapolation, we do not expect $n$ to exceed 15. Therefore for $n \leq 1000$ and the Lagrange multiplier
\begin{align}
    \mu = \frac{ -C_n(x_1-1)}{2\sin^2(\alpha) x_0 \gamma_0 (n+1)} 
\end{align}
we get
\begin{align}
 	C_n = -\frac{\mu}{n}\varphi_0 = \mu \varphi_{k\neq 0}.
\end{align}
This proves that Eq.\ \eqref{eq:sin2appendix} solves Eq.\ \eqref{eq:optimization} and is thus a candidate for a minimum or a maximum of the optimization problem (Eqs.\ \eqref{eq:optimization_real1}-\eqref{eq:optimization_real3}). Compared to other spacing methods, this choice has a lower $C_n$ such that it is reasonable to assume that the tilted Chebyshev nodes are indeed a minimum of the optimization problem. It should be noted that there might be other minima (that respect the ordering $x_0 < \dots <x_n$) leading to a lower minimum of $C_n$.

\section{Non-Markovian noise model \label{sec:AppendixNM}}
The non-Markovian noise model was derived using the following toy model. The system consists of a single qubit that is subjected to depolarization noise of strength $\lambda_\mathrm{M} = (1- \eta) \lambda$ and at the same time coupled to a second qubit, which simulates a highly non-Markovian environment, via the Hamiltonian $H = Z \otimes I + \lambda_\mathrm{NM} X \otimes X + I \otimes Z$. Here $\lambda_\mathrm{NM}= \eta \lambda$ is the non-Markovian coupling strength and $I$, $X$ and $Z$ denote the identity and the x- and z-Pauli matrices, respectively. The dynamics of the two qubit state $\rho_\lambda$ is given by the equation
\begin{align}
 	\frac{\partial}{\partial t} \rho_{\lambda}(t) = -i [H, \rho_{\lambda}] + \lambda_\mathrm{M}  \Big( \frac{\mathrm{tr}_2(\rho_{\lambda})}{2} - \rho_{\lambda} \Big). 
\end{align}
We used the initial condition $\rho_{\lambda_0}(0) = \frac{I+X}{2} \otimes \frac{I}{2}$. After time $\tau$ the expectation value $E(\tau)=\mathrm{tr}(\rho_{\lambda_0} A)$ of the observable $A = X \otimes I$ is given by
\begin{align}
 	E(t) = e^{-\tau \lambda_\mathrm{M}} &\Big( \cos(\tau \lambda_\mathrm{NM}) \cos(\tau \omega)  \notag \\
 	&+  \frac{\lambda_\mathrm{NM}}{\omega} \sin(\tau \lambda_\mathrm{NM}) \sin(\tau \omega) \Big)
\end{align}
where $\omega = \sqrt{4 + \lambda_\mathrm{NM}}$. For simplicity we chose $\tau=1$ which results in $E^* = \cos(2)$ and the noise model in Eq.\ \eqref{eq:nonMarkov}.

\section{Highly non-Markovian noise \label{sec:AppendixB}}
\begin{figure}[b]
	\centering
	\includegraphics[width=0.48\textwidth]{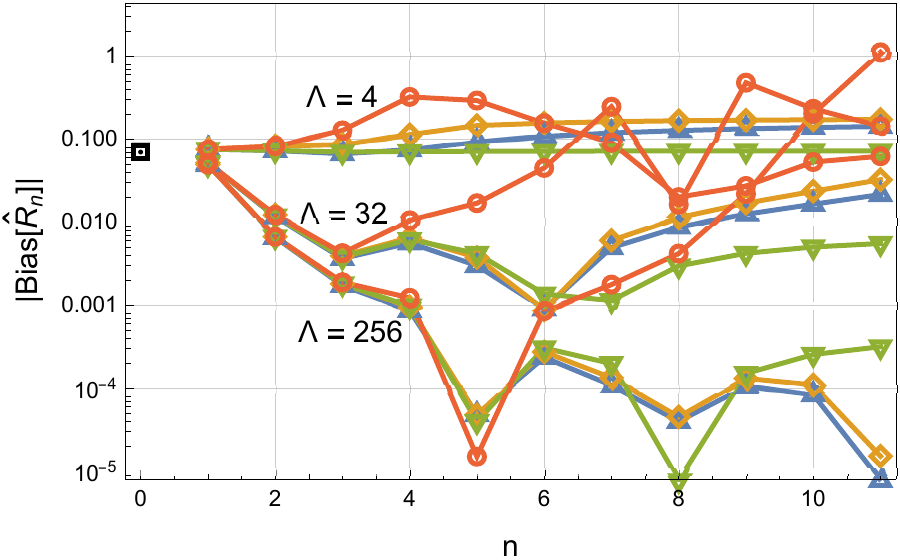}
	\caption{Comparison of the bias of Richardson extrapolation on the noise model in Eq.\ \eqref{eq:nonMarkov} with $\eta = 0.9$ and $\lambda_0 = 0.4$ for different $n$. The overhead in measurements is $\Lambda^2 = 4^2, 32^2$ and $256^2$, respectively. The different spacing methods are: tilted Chebyshev (blue, uptriangle), extremal Chebyshev (yellow, diamond), exponential (green, downtriangle) and equidistant (red, circle). The black square for $n=0$ is the bias without mitigation $|\mathrm{Bias}[\hat{R}_0]|$. \label{fig:NonMarkov_n09}}
\end{figure}

In the highly non-Markovian case, the mitigation behavior can be more complex than for purely Markovian noise. In Section \ref{sec:Experiments}, we demonstrated that in this case the mitigation result does not necessarily improve with larger $n$. Analogously to Figs.\ \ref{fig:Markov_n} and \ref{fig:NonMarkov_n02}, we plot the bias of $\hat{R}_n$ over $n$ in Fig.\ \ref{fig:NonMarkov_n09}, this time for a highly non-Markovian noise $\eta = 0.9$. 

For this noise function, the mitigation is more chaotic and a small change in one of the parameters $n$ or $\lambda_0$ can have a large influence on the mitigation result. We can see, that for small $\Lambda$, the mitigation result worsens for larger $n$. For $\Lambda=4$ this is the case once $n > 3$ and for $\Lambda=32$ for $n>6$. This confirms, that for a complex noise model, $n$ should not be chosen too large. From Fig.\ \ref{fig:Density_Grid} we can estimate that $\hat{n}$ for $\Lambda=4$, 32 and 256, are $\hat{n} = 1$, 2-3 and 5-6, respectively. For this specific noise model it seems, that $n \approx 2\hat{n}$ is a good choice for $n$. The parameter $\Lambda$, on the other hand, should still be chosen as large as possible. The larger $\Lambda$, the larger $n$ can be chosen and the more noise can be mitigated with Richardson extrapolation.

\end{appendix}

\end{document}